\documentclass[final, 12pt]{elsarticle}

\usepackage{graphicx}

\usepackage{amssymb}

\usepackage{lineno}
\usepackage{dcolumn}
\usepackage{bm,soul}

\usepackage[utf8]{inputenc}
\usepackage{mathptmx}
\usepackage{physics}
\usepackage{cancel}
\usepackage{verbatim}
\usepackage[colorlinks = true,
linkcolor = blue,
urlcolor  = blue, 
citecolor = blue,
anchorcolor = blue]{hyperref}
\usepackage{listings}
\usepackage[dvipsnames]{xcolor}

\definecolor{codegreen}{rgb}{0,0.6,0}
\definecolor{codegray}{rgb}{0.5,0.5,0.5}
\definecolor{codepurple}{rgb}{0.58,0,0.82}
\definecolor{backcolour}{rgb}{0.95,0.95,0.92}

\lstdefinestyle{mystyle}{
    commentstyle=\color{codegreen},
    keywordstyle=\color{blue},
    numberstyle=\tiny\color{codegray},
    stringstyle=\color{codepurple},
    basicstyle=\scriptsize,
    breakatwhitespace=false,
    breaklines=true,
    captionpos=b,
    keepspaces=true,
    showspaces=false,
    showstringspaces=false,
    showtabs=false,
    tabsize=2,
    language=Python
    }
\usepackage{float}

\usepackage{longtable}
\usepackage{makecell}

\usepackage{todonotes}

\newcounter{bla}

\journal{Computer Physics Communications}

\begin{document}

\begin{frontmatter}



\title{Sarkas: A Fast Pure-Python Molecular Dynamics Suite for Plasma Physics}


\author[label1]{Luciano G. Silvestri\corref{lux}}
\author[label1]{Lucas J. Stanek}
\author[label2]{Gautham Dharuman}
\author[label1]{Yongjun Choi}
\author[label1]{Michael S. Murillo}

\cortext[lux]{Corresponding author.\\\textit{E-mail address:} silves28@msu.edu}

\address[label1]{Michigan State University, East Lansing, MI, 48824}
\address[label2]{Lawrence Livermore National Laboratory, Livermore, CA, 94550}

\begin{abstract}
We present an open-source, performant, pure-python molecular dynamics (MD) suite for non-ideal plasmas. The code, Sarkas, aims to accelerate the research process by providing an MD code but also pre- and post-processing tools. Sarkas offers the ease of use of Python while employing the Numba library to obtain execution speeds comparable to that of compiled languages. The available tools in Sarkas include graphical displays of the equilibration process through a Jupyter interface and the ability to compute quantities such as, radial distribution functions, autocorrelation functions and Green-Kubo relations. Many force laws used to simulate plasmas are included in Sarkas, namely, pure Coulomb, Yukawa, and Moli\`ere pair-potentials. Sarkas also contains quantum statistical potentials and fast Ewald methods are included where necessary. An object-oriented approach allows for easy modification of Sarkas, such as adding new time integrators, boundary conditions and force laws. 
\end{abstract}
\begin{keyword}
Python, NumPy, Numba, Molecular Dynamics, Plasma.
\end{keyword}

\end{frontmatter}

\maketitle



{\bf PROGRAM SUMMARY}

\begin{small}
\noindent
{\em Program Title:} Sarkas \\
{\em CPC Library link to program files:} (to be added by Technical Editor) \\
{\em Developer's respository link:} https://github.com/murillo-group/sarkas \\
{\em Code Ocean capsule:} (to be added by Technical Editor)\\
{\em Licensing provisions:} MIT  \\
{\em Programming language:} Python \\
{\em Nature of problem:}
Molecular dynamics (MD) is an important tool for non-ideal plasma physics research. The wealth of MD codes available are not designed for plasma physics problems. The available codes are written in low-level languages and do not provide pre- and post-processing libraries. These are instead written by researchers in interpreted languages, forcing researchers to have a high level of computing background.\\
{\em Solution method:}
Development of a MD suite for plasma physics, complete of pre- and post- processing tools most commonly used in plasma physics. The suite is entirely written in Python for enhanced user-friendliness. The slow speed of Python is tackled by using the Numba library a just-in-time compiler for Python.

\end{small}

\section{Introduction}
\label{sec:intro}

Molecular dynamics (MD) is a powerful tool for simulating the microscopic dynamics of many-body systems. Many MD codes have been developed across scientific disciplines, including, for example, HOOMD for soft matter \cite{anderson2020hoomd}, GROMACS for biological applications \cite{abraham2015gromacs}, LAMMPS for materials modeling \cite{plimpton2007lammps} and VASP for electronic structure \cite{hafner2008ab}. In contrast, the field of computational plasma physics has traditionally relied on hydrodynamics \cite{larsen1994hyades,marinak2001three} and kinetic \cite{haack2017interfacial,larroche1993kinetic} codes because the temporal and spatial scales of interest are much too large for MD to be tractable; moreover, the detailed description of discrete particles is less important and instead macroscopic field variables are computed. However, MD plays a central role in a diverse set of plasma subfields concerned with understanding microscopic particles such as astrophysical systems \cite{horowitz2010crystallization,caplan2021precise}, dusty plasmas \cite{samsonov1999mach,li2019depinning}, ultracold neutral plasmas \cite{dharodi2020sculpted,pohl2004kinetic}, dense plasmas \cite{graziani2012large,murillo2008temperature}, low temperature plasmas \cite{graves2009molecular,hippler2008low}, plasma beams \cite{rahman1986structure,murphy2015increasing}, quark-gluon plasmas \cite{gelman2006classical,terranova2004constrained}, etc.

The earliest studies of plasmas using microscopic methods employed Monte Carlo methods \cite{Brush1966, Hooper1968}. The key computational issue for plasmas is handling the long range Coulomb interaction with an Ewald method \cite{hummer1995numerical}. Seminal work by Brush, Salin and Teller examined thermodynamic properties of the one-component plasma (OCP) in the mid-1960s \cite{Brush1966}. Hooper employed Monte Carlo methods for studies of the plasma microfield \cite{Hooper1968}. Substantial development of MD for plasmas was made in the subsequent decade. Development of the Particle-Particle-Particle-Mesh (PPPM) Ewald method allowed for MD simulations with as many as 10 000 particles as early as 1973 by Hockney, Goel and Eastwood \cite{Hockney1973}. Many contributions appeared by Hansen and McDonald, and their coworkers, to examine a wide range of physical phenomena \cite{Hansen1973,Pollock1973,Hansen1975,Hansen1979}. In particular, plasma-specific methods were developed for explicit-electron MD \cite{Hansen1978,Jones2007,feldmeier2000molecular}. Related charged systems, such as molten salts \cite{Hansen1975b,lantelme1974application} and liquid metals \cite{protopapas1973theory,mitra1978effective,tanaka1980molecular}, were studied in detail. More recent work has extended these methods to include electronic wave packet evolution \cite{klakow1994semiclassical}, momentum-dependent potentials \cite{dharuman2016atomic} and large-scale non-equilibrium MD with on-the-fly potentials \cite{stanton2018multiscale}. Very accurate interaction potentials for warm dense matter were recently obtained with force matching and from a neutral pseudoatom model \cite{stanek2021efficacy}. 

Historically, computational plasma physicists have developed their own codes; more recently, plasma physics models are being implemented in codes originally developed for other purposes. Such strategies require the plasma physicist to be part of the development process or have the skills and time to modify existing codes. Moreover, plasma physics observables are typically not available as post-processing packages. Here, we present Sarkas, a fast pure-python MD suite for plasma physics. Sarkas aims at lowering the entry barrier for computational plasma physics by providing a comprehensive MD suite complete with pre- and post-processing tools commonly found in plasma physics. It offers the ease of use of Python while being highly performant with execution speeds comparable to that of compiled languages. Its high-performance originates from the extensive use of \verb+NumPy+ arrays and \verb+Numba+'s just-in-time compilation. It offers a variety of interaction potentials commonly used in plasma physics. Furthermore, Sarkas' built-in pre-processing and post-processing libraries for data analysis allow researchers to get high-quality results with minimal effort.

\section{Code Structure and Capabilities}
\label{sec:code}

\begin{table}[ht]
\centering
\begin{tabular}{c | c}
\hline\hline
    Class  & Capabilities  \\
    \hline
    \texttt{Potentials} & Generalized $m-n$ Lennard-Jones potentials (LCL), \\ 
    & Exact Gradient Corrected Screening potential (LCL) \cite{Stanton2015}, \\ 
    & Moliere potential, \\
    & Yukawa (PPPM), \\ 
    & Coulomb (PPPM), \\ 
    & Quantum Statistical Potentials (PPPM)\cite{Jones2007}.\\
    \texttt{Integrator} & Velocity Verlet, \\
    & Magnetic Velocity Verlet \cite{Chin2008, Spreiter1999},\\
    & Magnetic Boris \cite{Chin2008},\\
    & Langevin integrator \\
    \texttt{Thermostat} & Berendesen  \cite{Berendsen1984}
    \\\hline
\end{tabular}
\caption{\label{tab:sim_table}Capabilities of each class. The terms (LCL) and (PPPM) indicate whether the potential is computed using the Linked-Cell-List (LCL) or Particle-Particle-Particle-Mesh (PPPM) algorithm.}
\end{table}

Sarkas aims at facilitating computational non-ideal plasma research by streamlining a researcher's workflow with a comprehensive infrastructure of fast and efficient libraries. Sarkas targets a broad user base: from experimentalists to computational physicists, from students approaching plasma physics for the first time to seasoned researchers. Therefore Sarkas' design revolves around two primary requirements: ease-of-use and extensibility. Sarkas is entirely written in Python without calls to \verb+C+ hence avoiding a two-language problem. It relies on the most common Python scientific packages, \textit{e.g.} \verb+NumPy+, \verb+Numba+, \verb+SciPy+, and \verb+Pandas+, which provide a solid foundation built, optimized, and well documented by one of the largest community of developers. Furthermore, Sarkas is developed using an object-oriented approach allowing users to add new features in a straight-forward way.

\begin{figure}[ht]
    \centering
    \includegraphics[width = \textwidth]{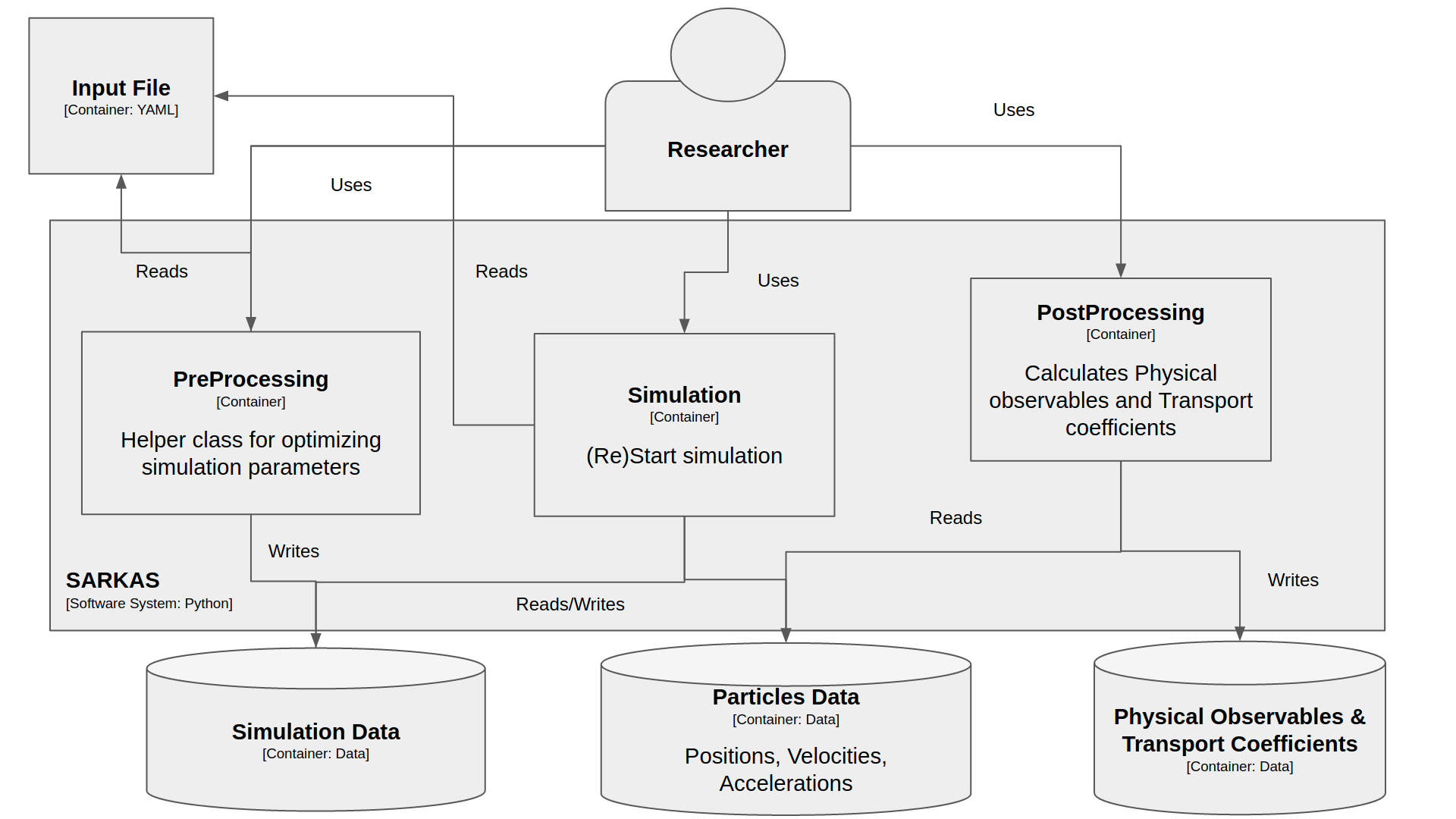}
    \caption{Diagram of Sarkas primary components and their relationships.}
    \label{fig:sarkas_diagram}
\end{figure}
Figure~\ref{fig:sarkas_diagram} shows a diagram of the main components of Sarkas. Users need only write an input file containing all the physical and simulation parameters of the system under investigation. The user interface is composed of three main classes: \verb+PreProcessing+, \verb+Simulation+, \verb+PostProcessing+.
Finally, the outputs are divided in three sets: Simulation data, Particles' data, and Physical Observables \& Transport Coefficients.
Using the \verb+PreProcessing+ class the researcher is able to optimize the parameters and update the input file. The  \verb+PreProcessing+ class saves to disk a set of files containing all the simulations data and class attributes \textit{e.g. physical constants and other parameters} needed for restarting simulations. The MD simulation is run using the \verb+Simulation+ class which evolves the system in time. Particles' trajectories, velocities, and accelerations are saved in compressed files at each timestep while thermodynamic quantities are saved in a \verb+CSV+ file. The desired physical quantities are computed via the \verb+PostProcessing+ class and saved in either \verb+CSV+ or \verb+HDF5+ files depending on the observable. 

The input file, written in \verb+YAML+, contains the necessary physical parameters of the system, \textit{e.g.} charge, density, temperature, number of species, and all the simulation parameters, such as timesteps and algorithm specific parameters. In addition, it allows users to provide a list of the physical observables to be calculated.

The three classes have two main methods: \verb+setup()+ and \verb+run()+. The \verb+setup()+ method, different than the usual \verb+__init()__+, is similar for all three classes and it handles the initialization of all the attributes of each class. The \verb+run()+ method accepts different inputs depending on the class, see below for more details. 

Each class is independent of the others allowing researchers to start and restart a project at any point. For example, once the input file has been finalized, a user is able to run multiple MD simulations, with different initial conditions as shown in the following script:
\begin{lstlisting}[language=Python, caption=Example of a script with comments for running multiple simulations and compute physical observables, label={lst:multi_run}]
# Import required libraries
import os
import numpy as np
from sarkas.processes import PreProcess, Simulation, PostProcess

input_file = os.path.join('input', 'ucp_N10k.yaml') # Path to input file
# Define a random number generator
rg = np.random.Generator( np.random.PCG64(154245) )

# Loop over the number of independent MD runs to perform
for i in range(5):
    seed = rg.integers(0, 15198)

    args = {
        'Parameters': {'rand_seed': seed}, # new rand_seed for each simulation
        'IO':   # Store all simulations' data in simulations_dir,
                # but save the dumps in different subfolders (job_dir)
            {
                'simulations_dir': 'UCP_DIH_N10k',
                'job_dir': 'run{}'.format(i)
            },
    }
    # Run the simulation.
    sim = Simulation(input_file)
    sim.setup(read_yaml=True, other_inputs=args)
    sim.run()
    # Calculate physical observables
    postproc = Postprocess(input_file)
    postproc.setup(read_yaml=True, other_inputs=args)
    postproc.run()
\end{lstlisting}
We note that the above scripts illustrates Sarkas' predisposition to data science as similar scripts can be run on clusters to produce large databases of plasma properties.

\begin{figure}[ht]
    \centering
    \includegraphics[width = \textwidth]{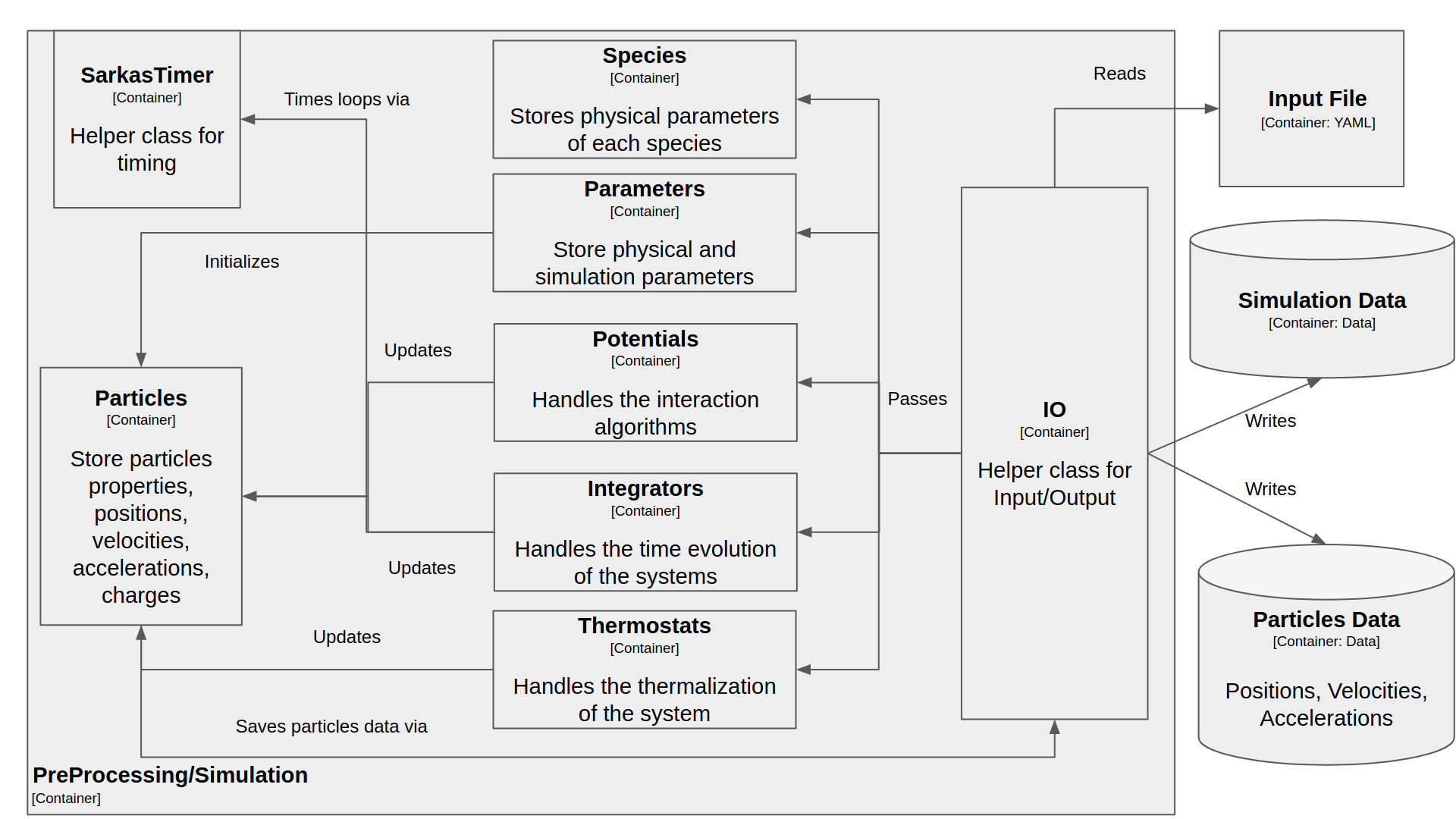}
    \caption{Diagram of the \texttt{PreProcessing} and \texttt{Simulation} classes. Each container is a class except for three containers outside the main class. Input File is a YAML file containing all the simulation parameters. Simulation Data and Particles Data indicate storage on disk of the relevant data.}
    \label{fig:sim_diagram}
\end{figure}
The structure of the \verb+PreProcessing+ and \verb+Simulation+ classes differ only in their hidden methods.  Figure~\ref{fig:sim_diagram} shows a diagram of the \verb+PreProcessing+ and \verb+Simulation+ classes, their component classes, and their relationships. More information about \verb+PreProcessing+ and \verb+PostProcessing+ will be given in Secs.~\ref{subsec:preproc} and \ref{sec:postproc}.

Each container in Fig.~\ref{fig:sim_diagram} represents a class that handles a particular task of the simulation stage. For example, the \verb+Integrator+ class handles the time integration of the simulation. It updates particle coordinates at each time step using one of the available integrators, see Tab.~\ref{tab:sim_table} for available integrators. The system is equilibrated via the \verb+Thermostat+ class using the Berendsen algorithm \cite{Berendsen1984}. The \verb+Potential+ class is used to calculate the total potential energy and force between particles and to update particles accelerations, see Tab.~\ref{tab:sim_table} for available potentials.
The first three are short range potentials and use a linked-cell-list (LCL) algorithm, while the last three are long range potentials that use a generalize PPPM algorithm presented in Ref.~\cite{Dharuman2017}. In particular, Sarkas is the only MD code, to our knowledge, that provides quantum statistical potentials and allows for the simulation of plasmas in a constant external magnetic field. Implementation of new integrators, thermostats, or potentials requires the addition of a new method to the corresponding class. 
The attributes of each class, except for \verb+Particles+, \verb+SarkasTimer+, and \verb+IO+, are saved as \verb+.pickle+ files (Simulation Data).

\begin{figure}
    \centering
    \includegraphics[width = \textwidth]{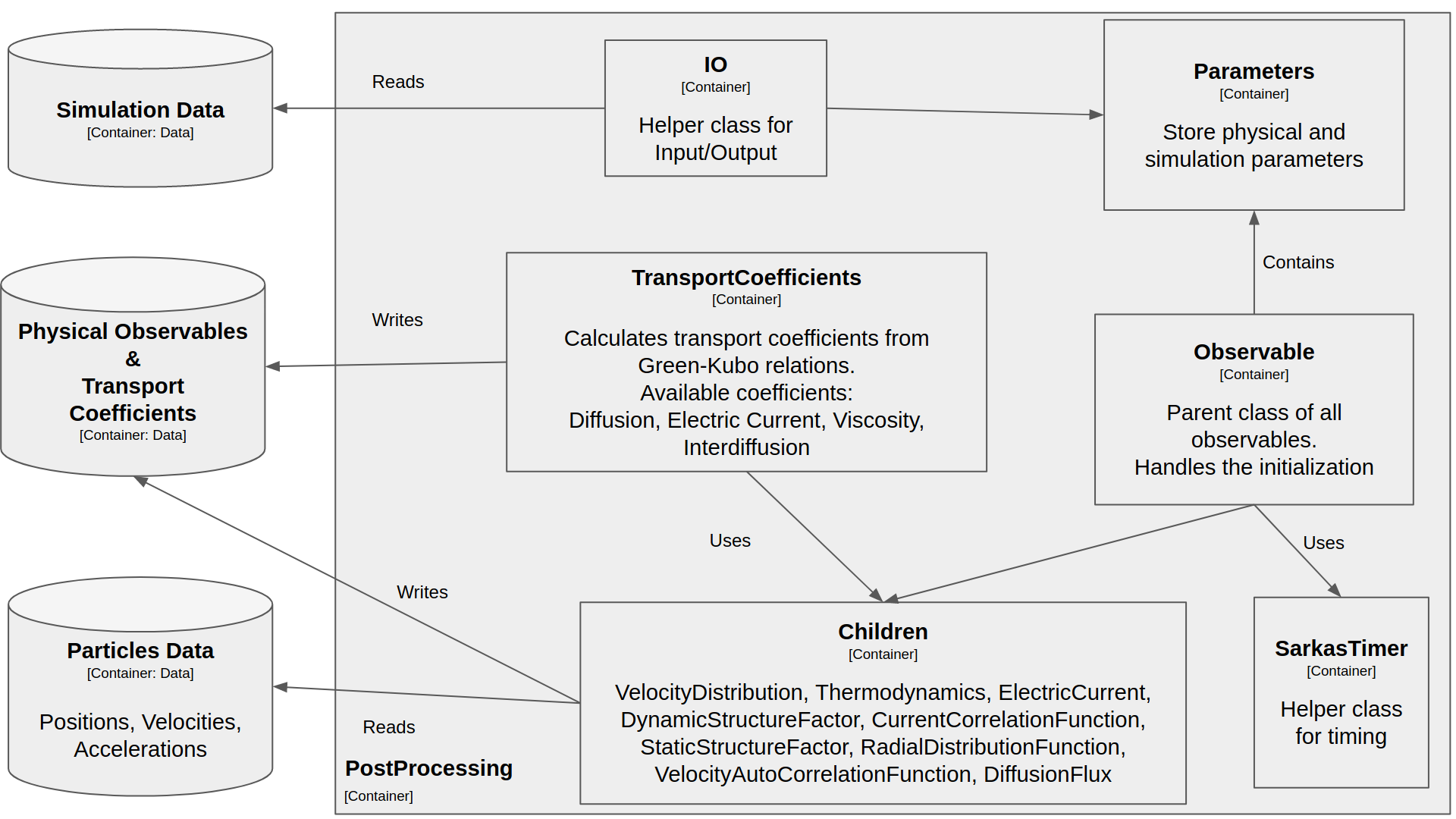}
    \caption{Diagram of the \texttt{PostProcessing} class.}
    \label{fig:postproc_diagram}
\end{figure}
Figure~\ref{fig:postproc_diagram} shows a diagram of the \verb+PostProcessing+ class and its component classes. The \verb+Observable+ is a parent class that handles the initialization and setup of each physical observables, represented as children classes. Each class reads in particles data and saves observables data to \verb+CSV+ or \verb+HDF5+ files. 
We note that besides the most common observables, \textit{e.g.} radial distribution function and the velocity autocorrelation function, Sarkas provides a class for the calculation of the velocity distributions, their moments, and the Hermite coefficients for the study of non-equilibrium systems such as ultracold neutral plasmas \cite{Sprenkle2020, Silvestri2021}.  

The most common transport coefficients are calculated by static methods of the \verb+TransportCoefficient+ class. Each method creates an instance of the required observable to calculate the autocorrelation function which is then integrated to obtain the desired transport coefficient. This is returned to the user as a \verb+Pandas.DataFrame()+ and saved to disk in \verb+CSV+ or \verb+HDF5+ files.

\section{Performance}
\label{sec:speed}
\begin{figure}[ht]
    \centering
    \includegraphics[width = 1.0\textwidth]{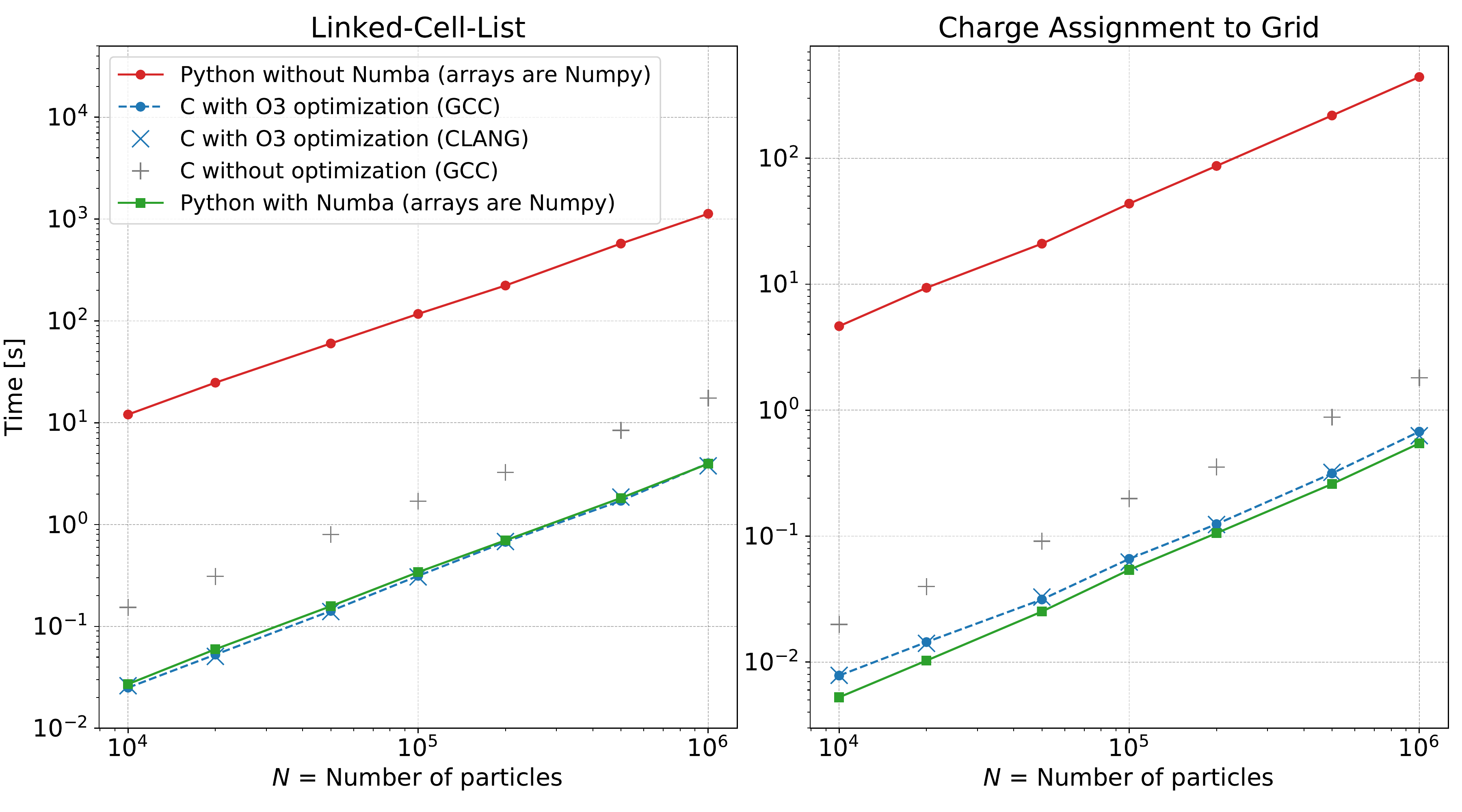}
    \caption{Mean execution times of LCL routine (left) and charge assignment routine (right) using \texttt{Numba} (squares) and without using \texttt{Numba} (red dots). Execution times of an equivalent code written entirely in \texttt{C} and compiled with \texttt{O3} optimization using \texttt{GCC} and \texttt{CLANG} compilers are shown in blue dots (dashed line) and blue crosses, respectively. Execution times for the \texttt{C} code without any compiler optimization (using \texttt{GCC}) are shown in grey plus signs. The standard deviation was found to be less than 3\%, too small to be visible in the plot. Computations performed on Intel Core i5-6360U CPU @ 2.00GHz and 8GB of RAM.}
    \label{fig:numba_timing}
\end{figure}
\begin{figure}[ht]
    \centering
    \includegraphics[width = 1.0\textwidth]{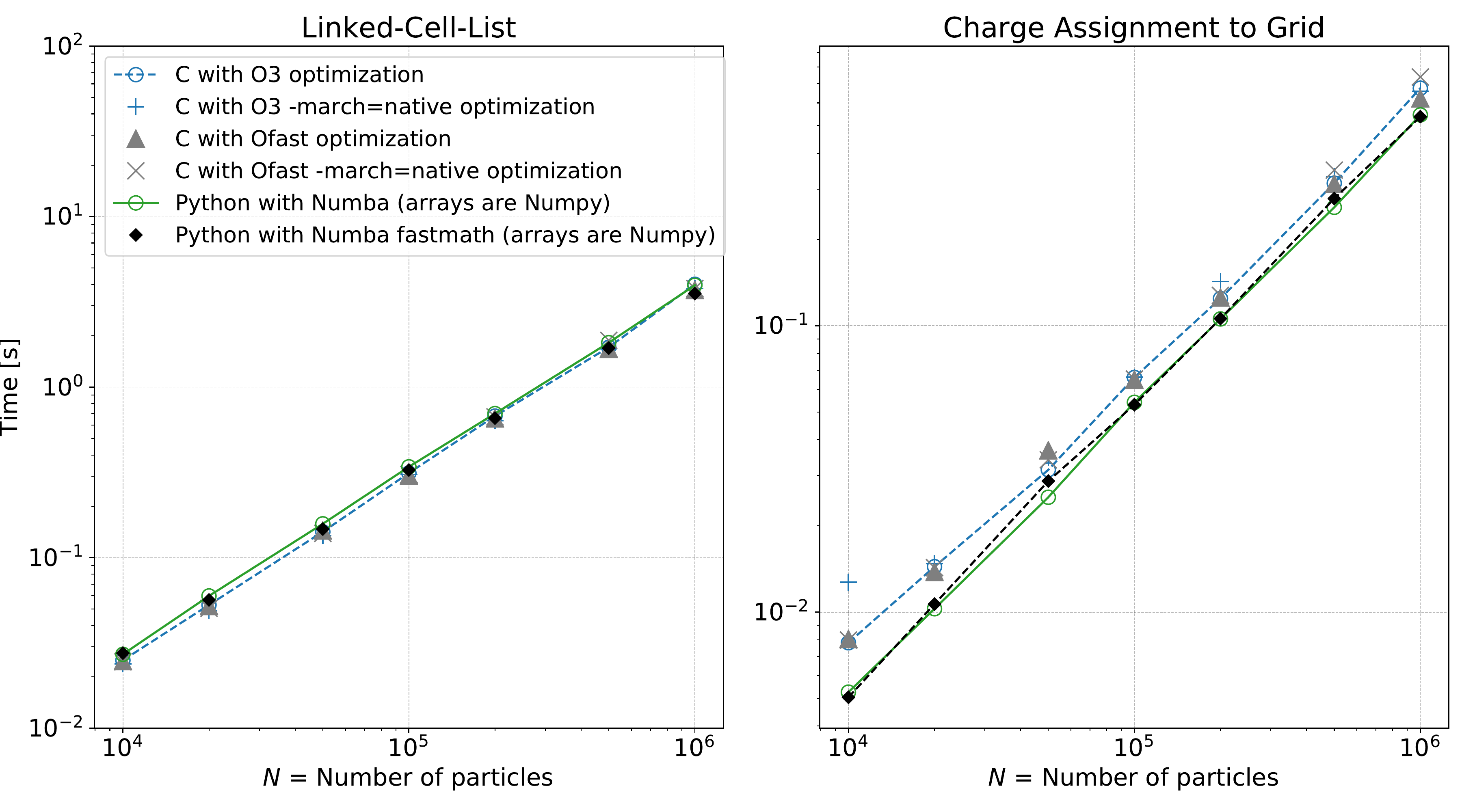}
    \caption{Mean execution times of LCL routine (left) and charge assignment routine (right) using \texttt{Numba} (green circles) and \texttt{Numba} with \texttt{fastmath} optimization (black diamonds) compared against C with different high performance compiler optimization (using \texttt{GCC}). Computations performed on Intel Core i5-6360U CPU @ 2.00GHz and 8GB of RAM.}
    \label{fig:numba_C_opt_timing}
\end{figure}

The speed bottleneck in a typical MD code is the force calculation which requires the calculation of the distance between all pairs of particles. The simplest algorithm scales as $O(N^2)$. Faster algorithms use linked cell lists (LCLs) and, in the case of long-range forces, Ewald summation is employed, usually within a mesh-based method \cite{Dharuman2017}. We recall that the former consists in dividing the simulation box into cells and then creating two linked lists. This algorithm scales as $O(N)$.
All of these requires several nested loops which are notoriously slow in Python due to the interpretive overhead. 
We circumvent this and other interpretative overheads by using \verb+Numba+, a just-in-time (JIT) compiler that translates a subset of Python and \verb+NumPy+ code into fast machine code \cite{Numba}. Implementation of Numba is straightforward and it requires the addition of a decorator around functions. Examples of the speed up and ease of use of \verb+Numba+ can be found in its documentation. 

In order to show Sarkas' performance we compare the execution times of two of its routines - the LCL routine and the routine for charge assignment to grid with an equivalent version of the code written entirely in \verb+C+. Both routines scale linearly with particle number, and therefore are some of the compute intensive parts of force calculation in Sarkas. The LCL is employed for short-range forces and in the particle-particle part of PPPM. The charge assignment routine is employed in the particle-mesh part of PPPM for mapping charges/fields from particles to grid and vice-versa. The other compute intensive part of PPPM is the FFT routine in Sarkas, but is not discussed in this section since it's a Python wrapper of the FFTW library whose performance is well known \cite{benchfftw}. 

We calculate the Coulomb potential for a system of $N$ charges at a constant density $n = 1.62 \times 10^{32} \; N / $m$^3$. The potential is cut off at a distance $r_c = 3.42808 \times 10^{-11}$ m. The mean and standard deviation of ten consecutive calculations are shown in Fig.~\ref{fig:numba_timing}.  We found the standard deviation to be less than 3\% in all cases. The green squares and red dots (with solid lines to guide the eye) indicate the execution times of the LCL and charge assignment routine with and without \verb+Numba+'s decorator \verb+numba.njit+, respectively. Execution times of an equivalent code written entirely in \verb+C+ and compiled with \verb+O3+ optimization using \verb+GCC+ and \verb+CLANG+ compilers are indicated in blue dots (with dashed line) and blue crosses, respectively. \verb+CLANG+ compiler was also a choice in this comparison since \verb+LLVM+s powering \verb+numba.njit+ are essentially \verb+CLANG+ based compilers. The optimized execution times using \verb+GCC+ and \verb+CLANG+ are found to be very similar for both the routines. Fig.~\ref{fig:numba_timing} also shows the execution times (grey plus) for the \verb+C+ code compiled with GCC, but without any compiler optimization. The plots show that the \verb+Numba+ optimized LCL code is $\sim300$ times faster than the Python version, $\sim5$ times faster than the \verb+C+ code without optimization, and is similar in performance to the \verb+C+ code with \verb+O3+ optimization. For the charge assignment routine, \verb+Numba+ optimized code is $\sim800$ times faster than the Python version, $\sim4$ times faster than the \verb+C+ code without optimization, and $\sim 1.5$ times faster than the \verb+C+ code with \verb+O3+ optimization. 

To further understand the performance gain due to \verb+Numba+, the \verb+C+ code was compiled with different high performance compiler optimizations (using \verb+GCC+). The mean execution times are shown in fig.~\ref{fig:numba_C_opt_timing} for LCL routine (left) and charge assignment routine (right). Also included are the execution times for \verb+Numba+ with \verb+fastmath+ optimization which is similar to the compiler optimization with \verb+Ofast+ flag for the \verb+C+ code. The LCL execution times using \verb+Numba+ with or without \verb+fastmath+ optimization are comparable to the execution times of the \verb+C+ code with the different compiler optimizations shown in Fig.~\ref{fig:numba_C_opt_timing}. For the charge assignment routine, the \verb+C+ code becomes comparable in performance to \verb+Numba+ when compiled using \verb+Ofast+ optimization and only for particle numbers $\sim$10$^6$. 

We point out that this analysis does not suggest that \verb+Numba+ leads to codes faster than \verb+C+. Instead, it is meant to show that a naive use of \verb+Numba+ yields performance comparable to \verb+C+ code with additional compiler optimizations. This indicates that the enhanced performance of \verb+Numba+ is potentially due to the JIT compilation performed by its \verb+LLVM+ compiler with high performance optimization flags.

\section{Pre Processing}
\label{subsec:preproc}
The main difference between plasmas and normal liquids is the long-range aspect of the Coulomb interaction between particles which requires the use of Ewald summation \cite{Hockney1981}. Sarkas uses the generalized PPPM algortihm presented in Ref.~\cite{Dharuman2017}. This algorithm requires nine parameters: the short-range cut-off ($r_c$), the Ewald screening parameter ($\alpha$), the number of mesh points per direction, the charge approximation order, and the number of aliases of the Fast Fourier Transform (FFT) per direction. MD simulations are often performed with sub-optimal parameters which lead to inefficient and longer runs. Furthermore, optimal parameters are not easily found as they depend not only on the type of problem under investigation, but also on the available computational hardware. Researchers are thus left to use trial-and-error approaches or rely on colleagues' suggestions to choose simulation's parameters. 
In the following we present useful metrics and an example use of the \verb+PreProcessing+ class for finding the optimal parameters. 

\subsection{Metrics}
Two important metrics are the force computation time, $\tau_F$, and the force error $\Delta F_{\rm tot}$. We estimate the force computation time as 
\begin{equation}
    \tau_F = a_0 + a_1 \left( \frac{4\pi n}{3} \frac{L^3}{N^3_c} \right ) N + a_2 +  5 a_3 M^3 \log_2(M^3),
    \label{eq:tau_f}
\end{equation}
where $a_i$ are hardware dependent computation times, $n$ is the number density of the plasma, $N$ the total number of particles, $L = L_x = L_y = L_z$ is the side of the simulation box, $N_c = L/r_c$ the number of cells of the LCL algorithm, and $M = M_x = M_y = M_z$ the number of mesh points per direction. The first two terms indicates the time of the particle-particle part while the third and fourth term indicate the approximate time for the particle-mesh part. We note that eq.~\eqref{eq:tau_f} is different than eq.~B1 of Ref.~\cite{Pollock1996}. The reason is twofold. First, we are interested in the relationship between computation time and $N_c$ and $M$ only, with a fixed number of particles $N$. The coefficients $a_0, a_2$, in fact, represent the time for creating the linked lists and assigning charges to the mesh, respectively, and they both scale as $O(N)$. Second, the FFT in Sarkas is performed via the FFTW library whose scaling is provided in their documentation page \cite{benchfftw}. 

The force error is defined as
\begin{equation}
    \Delta F_{\rm tot}(r_c, \alpha) = \sqrt{ \Delta F_{\rm PP}^2(r_c, \alpha) + \Delta F_{\rm PM}^2( \alpha)}
    \label{eq:delta_f_tot}
\end{equation}
where the subscripts indicate the error of the particle-particle (PP) or particle-mesh (PM) part of the calculation. Formulas for $\Delta F_{\rm PP} $ and $\Delta F_{\rm PM}$ depend on the type of interaction between particles. In the case of Yukawa potential, with inverse screening length $\kappa$, we have \cite{Dharuman2017}
\begin{equation}
    \Delta F_{\rm PP} = 2 \frac{(Ze)^2}{4\pi \epsilon_0} \sqrt{\frac{N}{V}} \frac{e^{-\alpha^2 r_c^2}}{\sqrt{r_c}} e^{-\kappa^2/4 \alpha^2},
    \label{eq:delta_f_pp}
\end{equation}
and 
\begin{equation}
    \Delta F_{\rm PM} =  \sqrt{\frac{N}{V}} \frac{(Ze)^2}{4\pi \epsilon_0} \frac{\chi}{\sqrt{V^{1/3}}}
    \label{eq:delta_f_pm}
\end{equation}
\begin{equation}
    \chi = \frac{1}{V^{1/3}} \left [ \left ( \sum_{\mathbf k \neq 0} G_{\mathbf k}^2 |\mathbf k |^2 \right )  -  \sum_{\mathbf n} \left [ \frac{\left ( \sum_{\mathbf m} \hat{U}_{\mathbf{k + m}}^2 G_{\mathbf{k+m}} \mathbf{k_n} \cdot \mathbf{k_{n + m}} \right )^2 }{ \left( \sum_{\mathbf m} \hat{U}_{\mathbf{k_{n+m}}}^2 \right )^2 |\mathbf{k_{n} }|^2 } \right ] \right ]^{1/2}.
\end{equation}
In the above equations $Z$ is the charge number, $e = |e|$ is the positive electron charge, $\epsilon_0$ is the vacuum permittivity, $V= L^3$ is the volume of the simulation box, and $G_{\mathbf k}$ is the optimal Green's function 
\begin{equation}
    G_{\mathbf k} = \frac{1}{\epsilon_0} \frac{ e^{-( \kappa^2 + \left |\mathbf k \right |^2)/(4\alpha^2)} }{\kappa^2 + |\mathbf {k}|^2}
\end{equation}
where $\mathbf k = \mathbf k_{\mathbf n}$ is the vector of allowed wavenumbers in the simulation box
\begin{equation}
    \mathbf{k_n} = \left ( \frac{2 \pi n_x}{L_x}, \frac{2 \pi n_y}{L_y},\frac{2 \pi n_z}{L_z} \right ), \quad n_{i} = - M_{i}, \dots, 0 \dots,  + M_{i}
\end{equation}
and $\hat{U}_{\mathbf k}$ is the Fourier transform of the B-spline of order $p$
\begin{equation}
    \hat U_{\mathbf{k}} = \left[ \frac{\sin(\pi n_x /M_x) }{ \pi n_x/M_x} \right ]^p
    \left[ \frac{\sin(\pi n_y /M_y) }{ \pi n_y/M_y} \right ]^p
    \left[ \frac{\sin(\pi n_z /M_z) }{ \pi n_z/M_z} \right ]^p,
\end{equation}
Finally $\mathbf{m}$ refers to the
triplet of grid indices $(m_x,m_y,m_z)$ that contribute to anti-aliasing. Note that in the above equations as $\kappa \rightarrow 0$, we recover the corresponding error estimate for the Coulomb potential. More details on the derivation of the equations can be found in the documentation and in Refs.~\cite{Dharuman2017,Kolafa1992, Stern2008} and references therein. 

There is no prescribed optimal value for the force error and it is up to users to decide. Inverting the above equations users can find optimal parameters $r_c,\; \alpha$ given some desired errors $\Delta F_{\rm {PP,PM}}$. However, while eq.~\eqref{eq:delta_f_pp} can be easily inverted for $r_c$, no simple formula is available for eq.~\eqref{eq:delta_f_pm} and users must calculate a Green's function for each chosen $\alpha$, a task that requires considerable computation time since $G_{\mathbf k}$ is a $M_x \times M_y \times M_z$ array. An analytical approximation of eq.~\eqref{eq:delta_f_pm} for the general case of nonzero $\kappa$ can be found for fast decaying $G_{\mathbf k}$ \cite{Dharuman2017},
\begin{equation}
    \Delta F_{\rm{PM}}^{(\textrm{approx})} \simeq \frac{(Ze)^2}{4\pi \epsilon_0} \sqrt{\frac{N}{V}} \left [ \frac{3}{2\pi^2} \sum_{m = 0}^{p -1 } C_{m}^{(p)} \left ( \frac{h}2 \right )^{2 (p + m)}\frac{2\beta(p,m)}{1 + 2(p + m)} \right ]^{1/2},
    \label{eq:pm_error_approx}
\end{equation}
\begin{equation}
    \beta(p,m) = \int_0^{\infty} dk \; \left ( \frac{1}{\epsilon_0} \frac{e^{-(\kappa^2 + k^2)/4\alpha^2}}{\kappa^2 + k^2} \right )^2 k^{2(p + m + 2)},
\end{equation}
where $h = L/M$ and the coefficients $C_m^{(p)}$ are listed in Table I of \cite{Deserno1998}. Despite not being invertible, eq.~\eqref{eq:pm_error_approx} is much faster to compute than eq.~\eqref{eq:delta_f_pm} and its use will be presented below.

We define a simulation to be efficient when the choice of the simulation parameters leads to a small $\Delta F_{\rm tot}$ and fast execution speeds. This suggests that a search for optimal parameters is not only a nine-dimensional minimization problem, but it also depends on the hardware and on the physical system under investigation. Furthermore, we are assuming that a small $\Delta F_{\rm tot}$ leads to more accurate particles' trajectories hence to more accurate postprocessed physical observables. However, we are not aware of any uncertainty quantification studies that support such statements nor on the effect of $\Delta F_{\rm tot}$ on the accuracy and precision of postprocessed quantities. 

In the following we present an example use of the \verb+PreProcessing+ class that allows users to find optimal parameters for their simulation. The search is limited to only the cut-off radius $r_c$ and Ewald screening parameter $\alpha$ as they are both present in Eqs.~\eqref{eq:tau_f}- \eqref{eq:delta_f_tot}.

\subsection{Example:}
The following is a typical script \footnote{Code ran on an Intel Core i7-8700K @ 3.70Ghz and 48GB of RAM running Ubuntu 18.04.} for running the \verb+PreProcessing+ class
\begin{lstlisting}[language=Python, caption=Example of Python script with comments, label={lst:preproc}]
# Import required libraries
import os
from sarkas.processes import PreProcess

input_file = 'yukawa_mks_p3m.yaml' # Define path to input file

preproc = PreProcess(input_file) # Initialize the class

preproc.setup(read_yaml=True) # Setup Sarkas classes all the parameters

preproc.run(
    timing=True, # time estimation
    loops = 20,   # the number of timesteps to average
    pppm_estimate=True, # Produce Force Error Plots
    timing_study = True) # produce time and force error maps
\end{lstlisting}

The file \verb+yukawa_mks_p3m.yaml+ contains the simulation and physical parameter of the system under consideration: a one component plasma with $N = 10\; 000$ ions with charge number $Z = 1$, mass $m = 1.673 \times 10^{-27}$ kg, at a density $n = 1.62\times 10^{32}$ N/m$^3$, temperature $T = 0.5$ eV, surrounded by an electron liquid with density $n_e = n$ and electron temperature $T_e = 1.25$ keV. The ions interact via a Yukawa potential of the form
\begin{equation}
     U(r) = \frac{Ze}{4\pi \epsilon_0 r}e^{-r/\lambda_{TF}},
\end{equation}
where $r$ is the distance between two particles and $\lambda_{\rm TF}$ is the Thomas-Fermi length given by the electron liquid properties. These parameters lead to $\Gamma = 101.23$ and $\kappa = a_{\rm ws}/\lambda_{\rm TF} = 0.5$. 
 
\begin{figure}[htp]
    \centering
    \includegraphics[width = \textwidth]{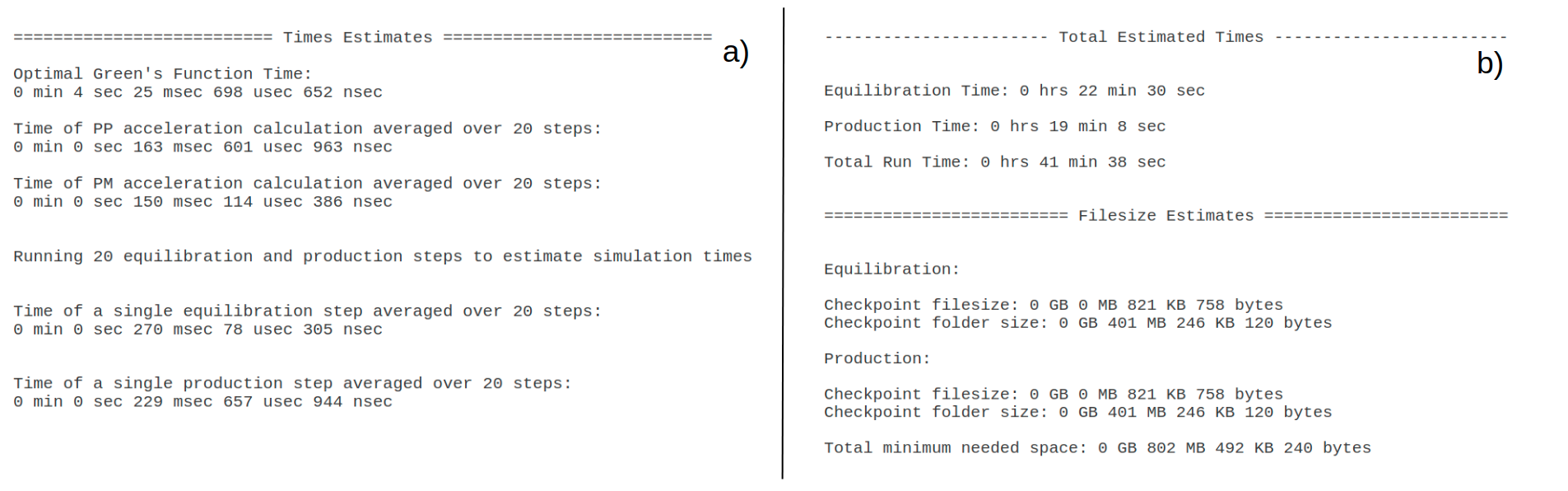}
    \caption{Section of the output of the script in Listing.~\ref{lst:preproc} split in two panels for better viewing. a) Timing of the acceleration and timesteps calculation. b) Total estimated times and size of the full simulation.}
    \label{fig:preproc_output}
\end{figure}

The script produces a verbose output to screen with a summary of all simulations parameters and relevant physical constants of the system. The option \verb+timing = True+ instructs to estimate the required space on disk and time of the full simulation by averaging the computation time of \verb+loops=20+ equilibration and production timesteps. The screen output of these option is shown in Fig.~\ref{fig:preproc_output}. We note that this information is of interest both for users running simulations on large clusters as well as personal desktops and/or laptops since it allows for better time and space management of hardware resources.

\begin{figure}[htp]
    \centering
    \includegraphics[width = \textwidth]{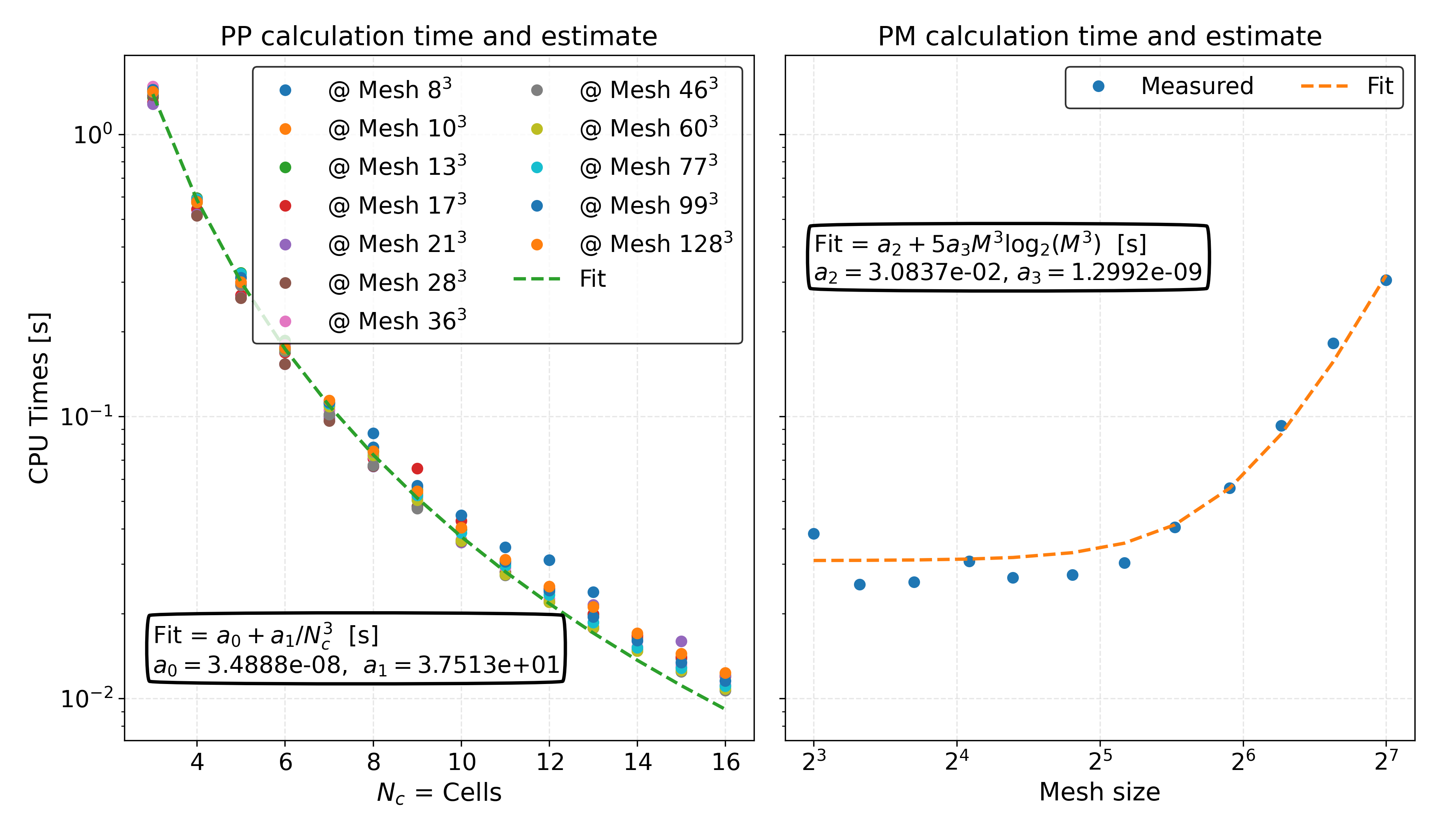}
    \caption{Time estimates of the PP and PM part of the force calculation with their respective fits.}
    \label{fig:time_fits}
\end{figure}

\begin{figure}[htp]
    \centering
    \includegraphics[width =\textwidth]{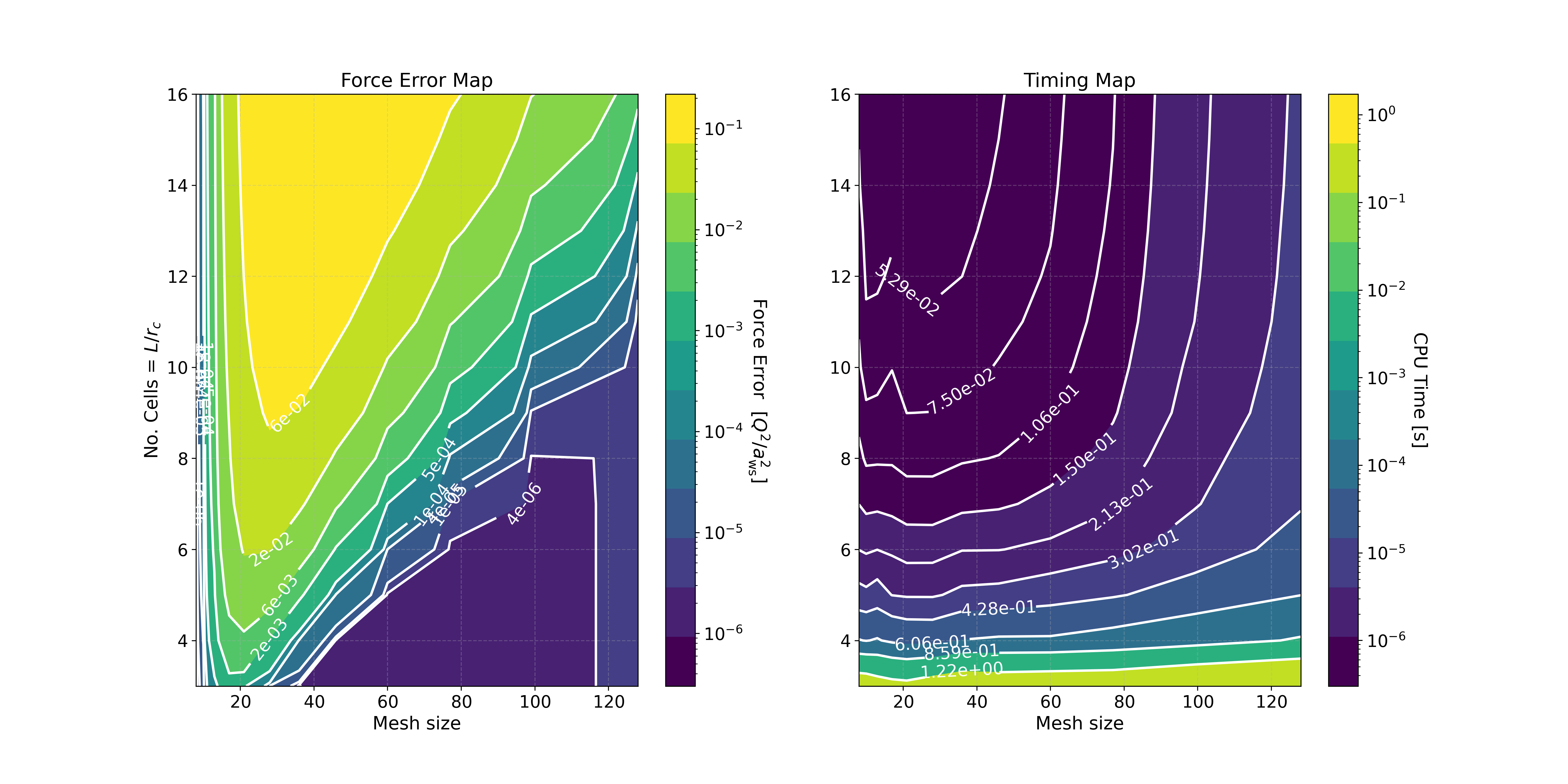}
    \caption{Contour maps of the total force error $\Delta F_{\rm tot}$ (left panel) and computation times (right panel). Colours indicates the magnitude of the force error $\Delta F_{\rm tot}$ of eq.~\eqref{eq:delta_f_tot} (left) and the total computation time as the sum of the data points in Fig.~\ref{fig:time_fits} (right)}
    \label{fig:tau_v_delta_F}
\end{figure}

The last two options are the primary advantage provided by Sarkas. The option \verb+timing_study = True+ is used to calculate the force error and estimate the $a_i$ coefficients of eq.~\eqref{eq:tau_f} by varying the number of cells $N_c$ and the mesh size $M$. This option produces the plots shown in Figs.~\ref{fig:time_fits} - \ref{fig:tau_v_delta_F}. Fig.~\ref{fig:time_fits} shows plots of the computation times of the PP and PM parts of the PPPM algorithm with their respective fits. The PP and PM part are fitted independently in order to show the correct scaling of each algorithm and to better identify the most time expensive between the two.

Note that for each $N_c$ the time of several mesh sizes is evaluated. This is because, while the computation time depends only on $r_c = L/N_c$, the force error $\Delta F_{\rm PP}$ depends also on the Ewald parameter, $\alpha$. There is no prescribed relation between $\alpha$ and $M$ as such Sarkas uses the heuristically chosen relation $\alpha = 0.3 M/L$. However, this choice is not particularly important as the objective is to calculate $\tau_F$ and to obtain an indicative value of $\Delta F_{\rm tot}$. The tuning of the $r_c$ and $\alpha$ parameters is obtained from the \verb+pppm_estimate+ option.

The total computation time, as the sum of the PP time and PM time, and the force error, calculated from eqs.~\eqref{eq:delta_f_tot}-\eqref{eq:delta_f_pm}, are then plotted as contour maps shown in Fig.~\ref{fig:tau_v_delta_F}.
Note that these maps are created using \verb+matplotlib.pyplot.contourf()+ method by passing the data points in Fig.~\ref{fig:time_fits} and not the fits, hence, the staggered contour lines. The fits are provided as a way for users to calculate the computation time for meshes and cells different than those computed by Sarkas. 
As expected the maps indicate that the smaller the force error the larger the computation time. They are meant to provide an overview of the parameter space and help users decide their optimal value for $M$ and approximate value of $r_c$. 
\begin{figure}[ht]
    \centering
    \includegraphics[width = \textwidth]{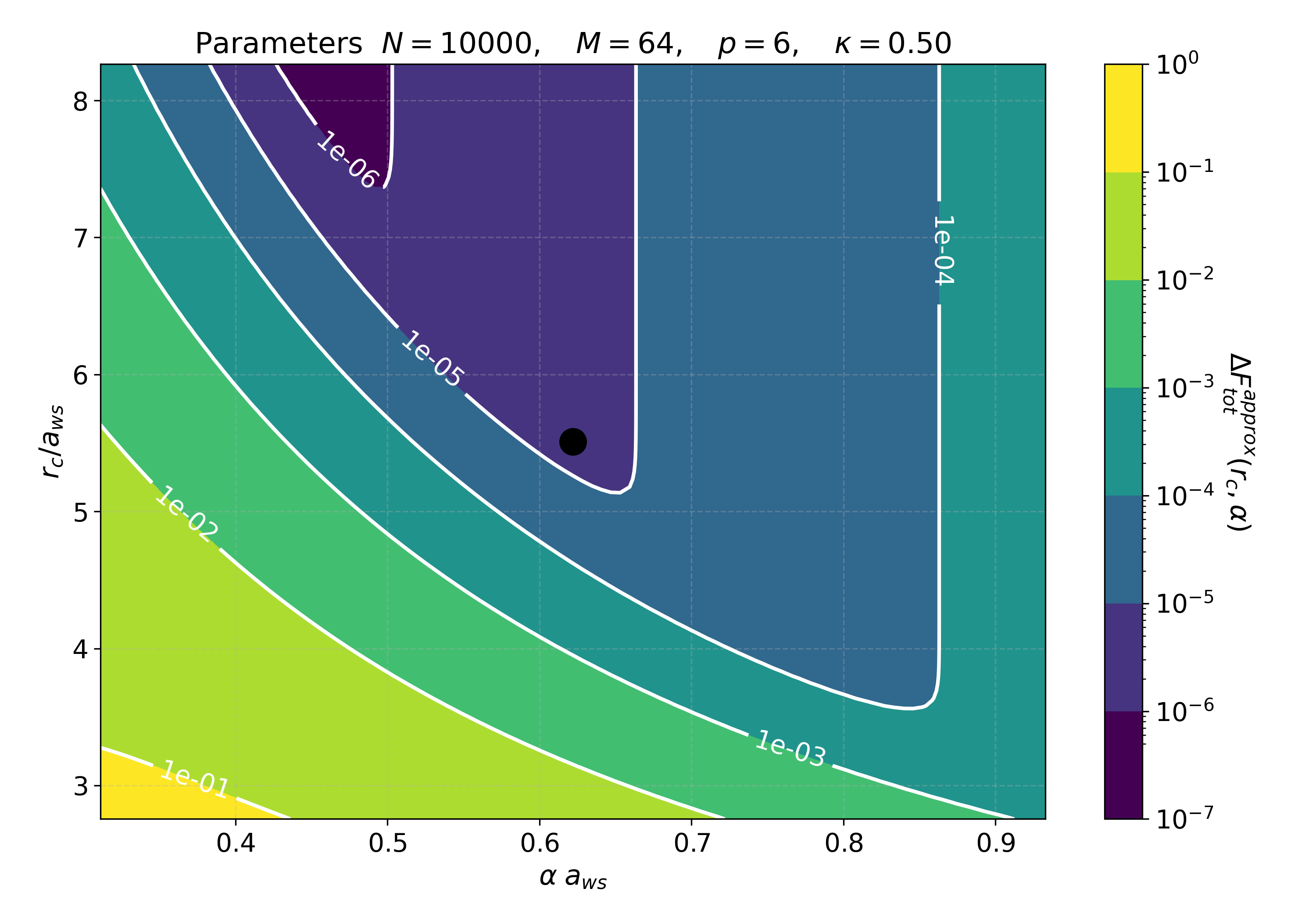}
    \caption{Contour map of approximate total force error $\Delta F_{\rm tot}^{\rm (approx)}$. The black dot indicates the original choice of $r_c$ and $\alpha$ as given in the input file.}
    \label{fig:ForceErr_clrmap}
\end{figure}

\begin{figure}[ht]
    \centering
    \includegraphics[width = \textwidth]{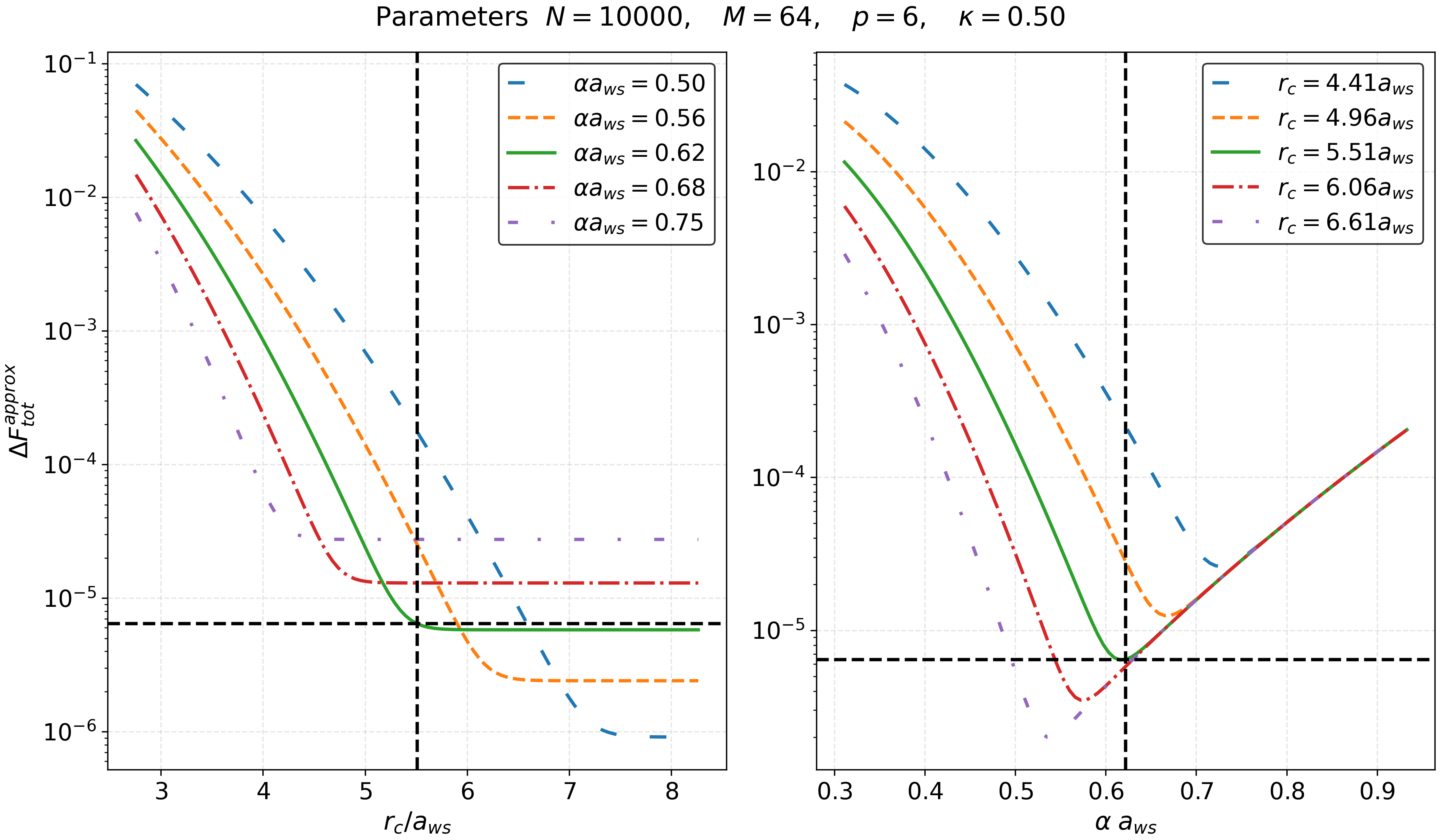}
    \caption{Line plots of approximate total force error $\Delta F_{\rm tot}^{\rm (approx)}(r_c, \alpha)$ as a function of $r_c$ at fixed values of $\alpha$ (left panel) and as a function of $\alpha$ at fixed values of $r_c$. The vertical dashed black line indicates the original choice of $r_c$ (left) and  $\alpha$ (right) as given in the input file. The horizontal black dashed line indicates the value of $\Delta F_{\rm tot}$ obtained from eq.~\eqref{eq:delta_f_tot}.}
    \label{fig:ForceErr_lineplot}
\end{figure}

The option \verb+pppm_estimate = True+ in Listing.~\ref{lst:preproc} is used for fine-tuning the values of $r_c$ and $\alpha$. This option produces two figures shown in Fig.~\ref{fig:ForceErr_clrmap} - \ref{fig:ForceErr_lineplot}. The first is a contour map in the ($r_c, \alpha$) parameters space of the approximate total force error
$\Delta F_{\textrm{tot}}^{(\textrm{approx})}$, see eq.~\eqref{eq:pm_error_approx}. 
The numbers on the white contours indicate the value of $\Delta F_{\textrm{tot}}^{(\textrm{approx})}$ along those lines and the black dot indicates the original choice of the $r_c$ and $\alpha$ parameters as provided in the input file. 

Note that this choice of parameters while being good, it might not be optimal. In order to find the best choice Sarkas provides Fig.~\ref{fig:ForceErr_lineplot}.
The left panel is a plot of $\Delta F_{\textrm{tot}}^{(\textrm{approx})}$ vs $r_c/a_{\rm ws}$ at
five different values of $\alpha a_{\rm ws}$ while the right panel is a plot of
$\Delta F_{\textrm{tot}}^{(\textrm{approx})}$ vs $\alpha a_{\rm ws}$ at five different values of $r_c/a_{\rm ws}$. The third line in both plots is evaluated at the original choice of $r_c,\alpha$. The vertical black dashed lines indicate the original choice of $\alpha a_{\rm ws}$ and $r_c/a_{\rm ws}$. The horizontal black dashed lines, instead, indicate the value of $\Delta F_{\rm tot}$ obtained from eq.~\eqref{eq:delta_f_tot}. 
These plots show that the analytical approximation of eq.~\eqref{eq:pm_error_approx} is a very good approximation and that the original choice of parameters is optimal as the intersection of the dashed lines falls exactly in the minimum of the curves.  
The left panel, in fact, indicates that larger values of $r_c$ lead to an inefficient code since it would calculate the interaction for many more particles without actually reducing the force error. Similarly, the right panel shows that larger values of $\alpha$ would increase the total force error.

\section{PostProcessing}
\label{sec:postproc}
The \verb+PostProcessing+ class is the most important class and what further separates Sarkas from other MD codes. This class handles the calculation of the desired physical observables. We note that other post-processing codes are available \cite{mdanalysis2011, Jamali2019}, but are often independent of the code used to perform the MD simulation; this requires users to install additional software and learn another application programming interface. 

\subsection{Example 2: Interdiffusion in a Binary Ionic Mixture}


As part of the \verb+PostProcessing+ class, Sarkas has the capability to compute any generic auto-correlation function \cite{HansenBook2013}. These auto-correlation functions have particular importance in the context of particle transport via Green-Kubo relations. We highlight this capability by calculating the interdiffusion coefficient a H-He Binary Ionic Mixture (BIM) \cite{HANSEN1985472}. Results reported in this section were performed on a 2.7 GHz Quad-Core Intel Core i7 processor with 16 GB 2133 MHz LPDDR3 memory, with macOS Catalina version 10.15.7. The auto-correlation function of importance for computing the interdiffusion coefficient has the form
\begin{equation}
    \rm{ACF_{ID}} = \langle \mathbf{j}(0) \cdot \mathbf{j}(t) \rangle,\label{eq:ACF}
\end{equation}
Where $\mathbf{j}(t)$ is the interdiffusion current defined as
\begin{equation}
    \mathbf{j}(t) = x_2\sum_{j=1}^{N_1} \mathbf{v}_{1,j}(t) - x_1\sum_{j=1}^{N_2}\mathbf{v}_{2,j}(t).\label{eq:IDC}
\end{equation}
where, $N_i$ is the number of species $i$, $N = N_1 + N_2$, $x_i = N_i/N$ is the concentration of species $i$, and $\mathbf{v}_{i,j}$ is the velocity of the $j$th particle of species $i$. The evolution of the interdiffusion current is assumed to be stationary and include all interparticle interactions. For molecular dynamics simulation, the ensemble average for an observable $\mathcal{O}$ is defined in terms of discrete time steps $\Delta t$ with phase space vectors $x_{n\Delta t}$ for $M$ total states
\begin{equation}
    \langle \mathcal{O} \rangle = \frac{1}{M}\sum_{n=0}^{M}\mathcal{O}(x_{n\Delta t}).
\end{equation}

The interdiffusion coefficient is calculated by the Green-Kubo relation \cite{zhou1996green}

 
\begin{equation}
\label{eq:GK}
    D_{12} = \frac{\mathcal{J}}{3Nx_1x_2} \int_0^\infty dt {\rm{{ACF}_{ID}}} = \frac{\mathcal{J}}{3Nx_1x_2} \int_0^\infty dt \langle \mathbf{j}(0) \cdot \mathbf{j}(t) \rangle,
\end{equation}
where the prefactor $\mathcal{J}$ in (\ref{eq:GK}) is the thermodynamic factor 
\begin{equation}
    \mathcal{J} = \frac{x_1x_2}{S_{cc}(k = 0)},\label{eq:ThF}
\end{equation}
where $S_{cc}(k)$ is the concentration-concentration structure factor that can be decomposed into partial structure factors as 
\begin{align}
\label{eq:scc}
    S_{cc}(k) = x_1x_2[x_2S_{11}(k) + x_1S_{22}(k) - 2\sqrt{x_1x_2}S_{12}(k)].
\end{align}
Calculating the interdiffusion coefficient (\ref{eq:GK}) is two-fold in that we must compute the auto-correlation function (\ref{eq:ACF}) and the thermodynamic factor (\ref{eq:ThF}). 

\begin{figure}
    \centering
    \includegraphics[width=\textwidth]{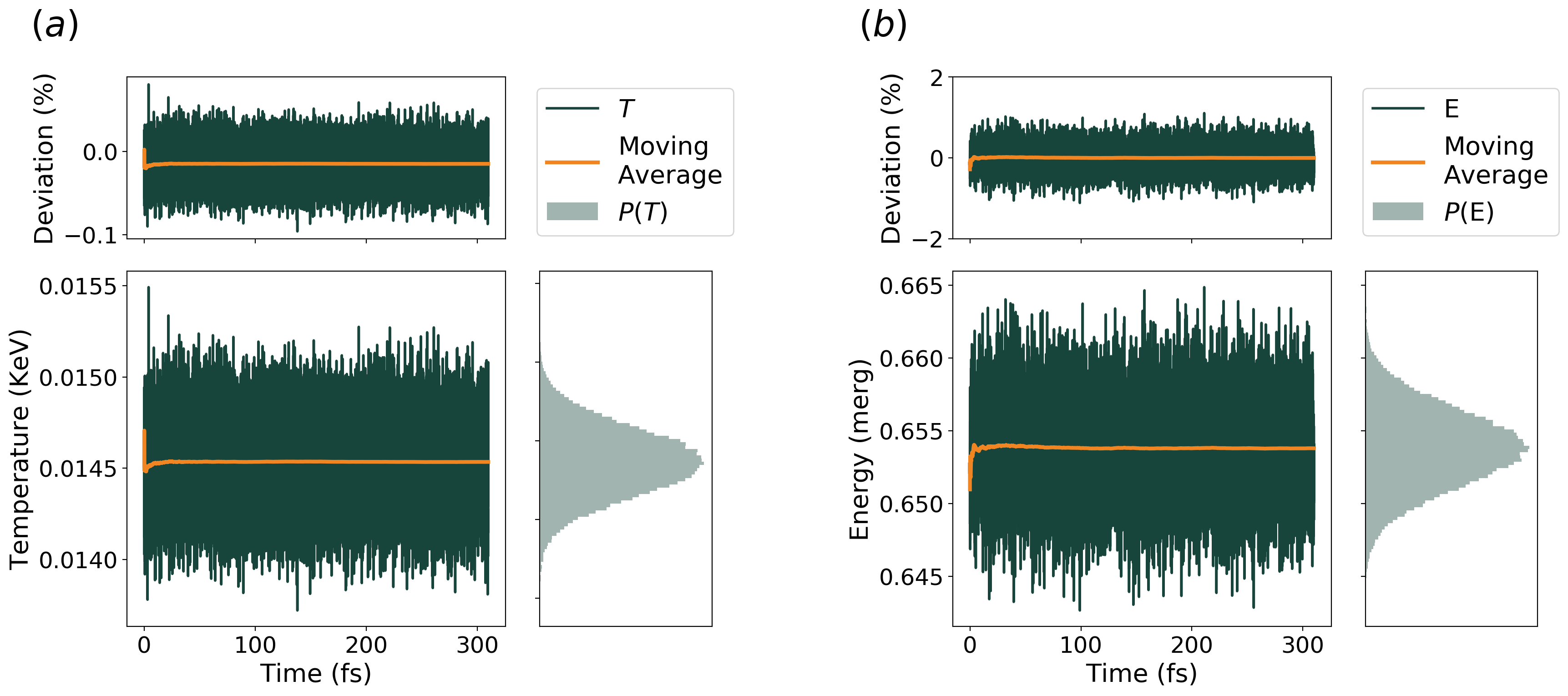}
    \caption{System information collected in the NVE ensemble for a H-He BIM with $\Gamma_{ij} = 40$ \cite{HANSEN1985472}. $(a)$ the temperature of the system. The temperature deviation is computed from the desired temperature of the system. $(b)$ the total energy of the system. The top plot shows the deviation from the average total energy. This plot aims to inform the user of possible equilibrium issues by showing moving averages of temperature and energy versus simulation time. The density plots show also highlight potential temperature and energy drifts by skewing the distributions.}
    \label{fig:eq_pr_check}
\end{figure}

We proceed by performing a molecular dynamics simulation for a H-He BIM with conditions described in Table~\ref{tab:H-He_gamma40}. We compare our results to those reported by Hansen et al. \cite{HANSEN1985472}. To check temperature or energy drift did not occur in the equilibration or production phases of the simulation, we plot the temperature and energy versus time in Fig.~\ref{fig:eq_pr_check}. We see that the deviation from the desired temperature is less than 0.1\% and no energy drift occurred. In the event a drift did occur, the distributions to the right of columns $(a)$ and $(b)$ would be skewed. A similar plot to Fig.~\ref{fig:eq_pr_check} can be generated with Sarkas' \verb+PostProcessing+ class using the code in Listing~\ref{lst:thermo}. 

\begin{lstlisting}[language=Python, caption={Generate plot of temperature and energy versus time, allowing for a quick analysis of the system's thermodynamic properties.}, label={lst:thermo}]
    from sarkas.processes import PostProcessing
    
    # Specify input file path
    postproc = PostProcessing(input_file)
    postproc.setup(read_yaml = True)
    
    # Equilibration phase check
    postproc.therm.setup(postproc.parameters)
    postproc.therm.temp_energy_plot(postproc, phase = 'equilibration')
    
    # Production phase check
    postproc.therm.temp_energy_plot(postproc, phase = 'production')
\end{lstlisting}

\begin{figure}
    \centering
    \includegraphics[width=\textwidth]{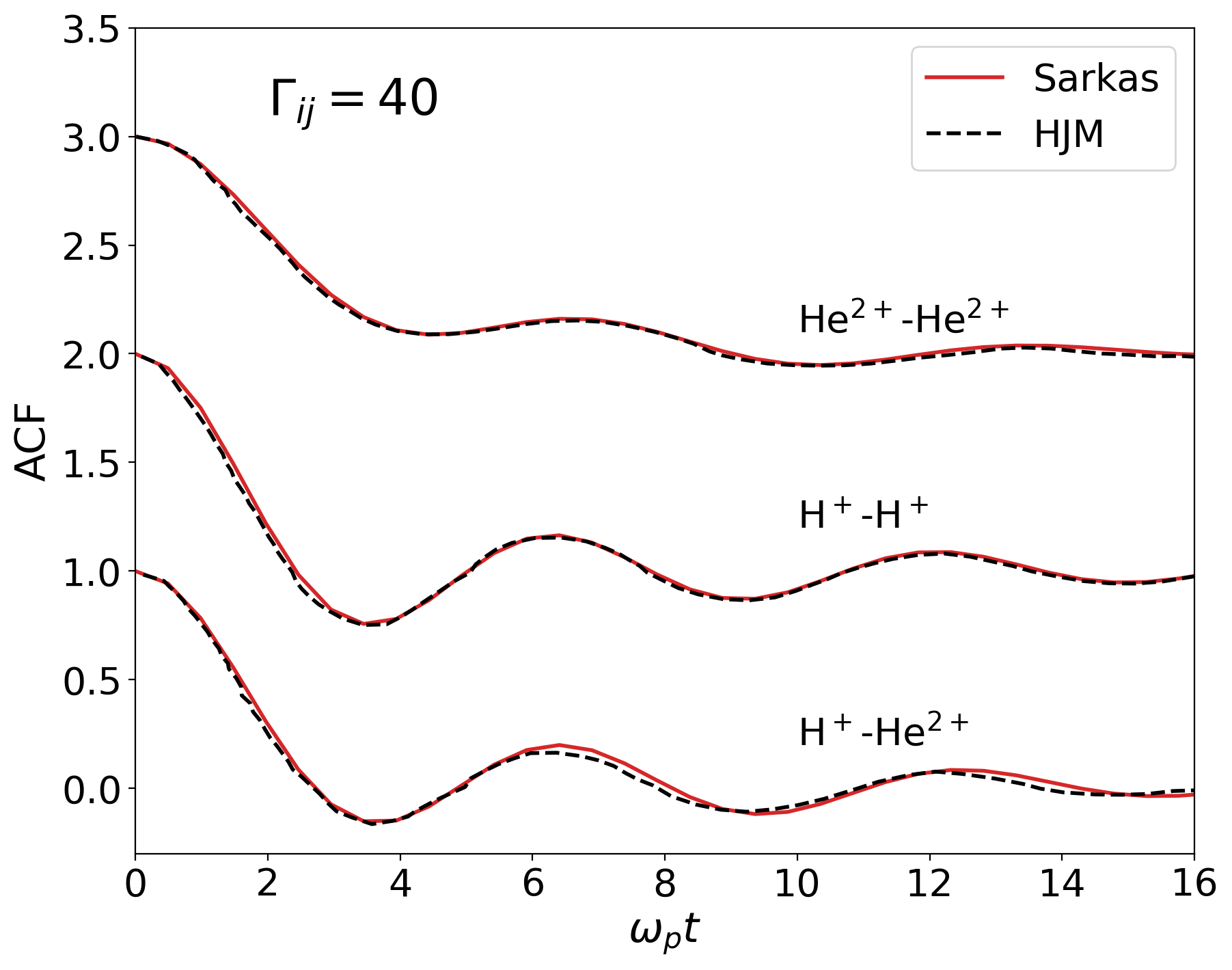}
    \caption{Comparison of auto-correlation functions for a H-He BIM mixture against results from \cite{HANSEN1985472}. Each autocorrelation function is an average of four.}
    \label{fig:HJM_ACF_compare}
\end{figure}

After verifying that the MD simulation is acceptable and no temperature or energy drift has occurred, we shift our focus to generating the interdiffusion auto-correlation function. In Sarkas, the interdiffusion auto-correlation function is computed as described in Listing~\ref{lst:lukecode}. As (\ref{eq:ACF}) is subject to statistical noise from small simulations, we compute the average of four autocorrelation functions of equal length. The averaging of correlation functions is carried out in Sarkas by specifying the \verb+no_slices=4+ variable. The value ``4" denotes that the simulation data is broken up into four equal consecutive chunks and an autocorrelation function is computed for each chunk. The autocorrelation functions using the above approach are shown in Fig.~\ref{fig:HJM_ACF_compare} where we compare our results to those obtained from \cite{HANSEN1985472}.

\begin{lstlisting}[language=Python, caption={Compute the autocorrelation functions for an H$^+$-He$^{2+}$ BIM.}, label={lst:lukecode}]

# Import the necessary methods for calculating the autocorrrelation functions
from sarkas.tools.transport import TransportCoefficient
from sarkas.tools.observables import VelocityAutoCorrelationFunction, DiffusionFlux

# Compute the interdiffusion auto-correlation function 
jc_acf = DiffusionFlux()
jc_acf.no_slices = 4 # Slice data into 4 chunks 
jc_acf.setup(postproc.parameters)
jc_acf.compute() 

# Similarly for the single species velocity autocorrelation function
vacf = VelocityAutoCorrelationFunction()
vacf.setup(postproc.parameters)
vacf.compute(no_slices = 4)

\end{lstlisting}

\begin{table}
\centering
\begin{tabular}{c r r}
\hline\hline
    Parameter  & This work & Ref.~\cite{HANSEN1985472} \\\hline
    $N_{part}$ & 2000 & 250\\
    $x_1$      & 0.5 & 0.5 \\
    $N_{t}$    & 100,000 & 5,000\\
    $dt^*$     & 0.49 & 0.15\\
    $\Gamma_{ij}$ & 40.01 & 39.74 \\\hline
\end{tabular}
\caption{\label{tab:H-He_gamma40}Simulation parameters for an H$^+$-He$^{2+}$ BIM. Note that $N_{part}$ denotes the total number of particle, $N_t$ denotes the number of time steps, $dt^* = dt/{\omega_pa_i}$ is the normalized time step as given in \cite{HANSEN1985472} and $\Gamma_{ij}$ is the cross-species Coulomb coupling parameter (see \cite{HANSEN1985472}). These simulation parameters were chosen to compare to \cite{HANSEN1985472}.}
\end{table}


Sarkas has the ability to integrate the resulting autocorrelation function (\ref{eq:ACF}) as demonstrated in Listing~\ref{lst:int_acf}. After computing the integral of \eqref{eq:ACF} it only remains to compute eq.~\eqref{eq:scc} which can also be done using Sarkas. The core to compute the partial structure factors is given in Listing~\ref{lst:Sk}. Note that (\ref{eq:scc}) must be evaluated for $k=0$. To evaluate (\ref{eq:scc}) for $k=0$, a large simulation cell is necessary and if not, an extrapolation to the $k=0$ limit must be performed. This extrapolation is left to the user as it is problem dependent. 

\begin{lstlisting}[language=Python, caption={Compute the integral of the interdiffusion autocorrelation functions for an H$^+$-He$^{2+}$ BIM.}, label={lst:int_acf}]
    # Integral of the auto-correlation function versus time
    int_acf = TransportCoefficient.interdiffusion(postproc.parameters, no_slices=4)
\end{lstlisting}

\begin{lstlisting}[language=Python, caption={Compute the partial structure factors of an H$^+$-He$^{2+}$ BIM.}, label={lst:Sk}]
# Compute the partial structure factors
postproc.ssf.angle_averaging = 'full' # Radial average for improved statistics
postproc.ssf.setup(postproc.parameters)
postproc.ssf.compute()
\end{lstlisting}

\begin{figure}
    \centering
    \includegraphics[width=\textwidth]{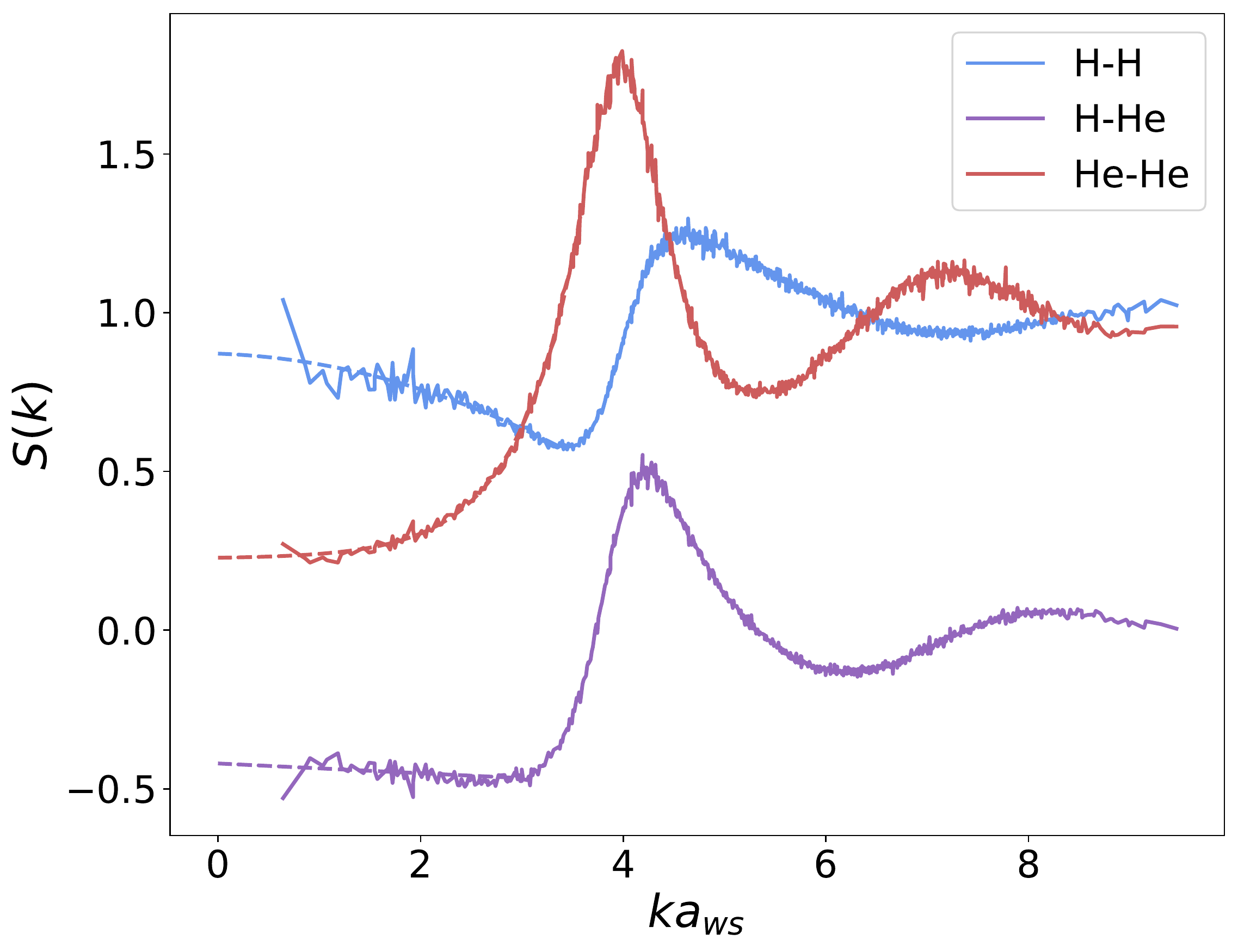}
    \caption{The static structure factors computed with the post processing tools in Sarkas for a H$^+$-He$^{2+}$ BIM for $\Gamma_{ij} = 40$. The dashed line approaching $k = 0$ denotes an extrapolation in order to compute the thermodynamic factor using (\ref{eq:scc}). In this case, we computed $\mathcal{J}= 1.03$.}
    \label{fig:my_label}
\end{figure}

\section{Conclusions and Outlook}
\label{sec:conclusions}
In conclusion, we have developed an open-source comprehensive MD suite for the simulation and analysis of non-ideal plasmas. Sarkas novelty lies in the availability of a preprocessing library, to aid researchers in running efficient simulations, and a post-processing library, for the calculation of the most common physical observables. Sarkas is entirely written in \verb+Python+ because of its high level of user-friendliness. We have shown that this choice does not harm the performance of the code as it compares to equivalent code written in \verb+C+. In order to demonstrates Sarkas ease-of-use, we have provided example scripts using the \verb+PreProcessing+ and \verb+PostProcessing+ classes.

Sarkas was developed for the needs of computational, theoretical, and experimental plasma physicists and for users with different levels of computing knowledge. Beginner users are able to optimize a simulation, run a simulation, and analyze data by providing a input file and a script with less than 20 lines of code, see documentation link for an example. Intermediate and advanced users are able to easily implement new features and extend Sarkas capabilities to their desires. 
Furthermore, Sarkas' documentation provides a growing list of examples usage for the most common plasma physics problems; from the classical one-component plasma model and its magnetized case to the study of non-equilibrium ultracold neutral plasmas. 

The authors would like to acknowledge funding from the Air Force Office of Scientific Research (AFOSR) Grant No.~FA9550-17-1-0394. Gautham Dharuman's work was performed in part under the auspices of the U.S. Department of Energy by Lawrence Livermore National Laboratory under Contract DE-AC52-07NA27344. All the codes and figures presented in this paper are available in Sarkas' documentation at https://murillo-group.github.io/sarkas.

\end{document}